\documentclass[pdftex]{nature}
\usepackage{graphicx}
\usepackage{siunitx}
\usepackage{amsfonts}
\usepackage{amsmath}

\newcommand\blfootnote[1]{%
  \begingroup
  \renewcommand\thefootnote{}\footnote{#1}%
  \addtocounter{footnote}{-1}%
  \endgroup
}

\title{Multimode laser cooling and ultra-high sensitivity force sensing with nanowires}
\author{Mahdi Hosseini$^{\dag}$, Giovanni Guccione, Harry J.\ Slatyer, Ben C.\ Buchler and Ping Koy Lam$^*$}

\begin{document}

\maketitle

\begin{affiliations}
\item{Centre for Quantum Computation and Communication Technology, Department of Quantum Science, Research School of Physics and Engineering, The Australian National University, Canberra, ACT 0200, Australia}
\end{affiliations}

\begin{abstract}
Photo-induced forces can be used to manipulate and cool the mechanical motion of oscillators. When the oscillator is used as a force sensor, such as in atomic force microscopy, active feedback is an enticing route to enhancing measurement performance. Here, we show broadband multimode cooling of $\mathbf{-23}$~\SI{}{\textbf{dB}} down to a temperature of $\mathbf{8 \pm 1}$~\SI{}{\textbf{K}} in the stationary regime. Through the use of periodic quiescence feedback cooling, we show improved signal-to-noise ratios for the measurement of transient signals. We compare the performance of real feedback to numerical post-processing of data and show that both methods produce similar improvements to the signal-to-noise ratio of force measurements. We achieved a room temperature force measurement sensitivity of $\mathbf{< 2\times10^{-16}}$~\SI{}{\textbf{N}} with integration time of less than $\mathbf{0.1}$~\SI{}{\textbf{ms}}. The high precision and fast force microscopy results presented will potentially benefit applications in biosensing, molecular metrology, subsurface imaging and accelerometry.
 \end{abstract}

\maketitle

\blfootnote{$^{\dag}$ New address: Massachusetts Institute of Technology, MIT-Harvard Center for Ultra Cold Atoms, Cambridge, MA 02139}
\blfootnote{$^*$ Corresponding Author email: ping.lam@anu.edu.au}

\section*{Introduction}
To attain high-resolution mass, position and force sensing, cantilevers with low effective masses and high mechanical quality factors are required. Nano-scale resonators are excellent candidates for mass and force sensing~\cite{Yazd:Lang:08, Zhao:IOP:2010, Arcizet:PRL:2006}, biomaterial sensing by means of Kelvin probe force microscopy~\cite{Gao:ABC:2009}, and single-spin and charge detection~\cite{Rugar:Nat:04}. Miniaturized oscillators, however, are more susceptible to thermal noise and their use can be challenging as it entails low-noise high-frequency electronics to monitor and control the vibrational modes. Passive cooling is not always an option, particularly in biomaterial sensing applications where samples are dependent on specific environmental conditions. Active feedback cooling can instead be used to reduce the Brownian motion of the oscillators. 

To date, effective optical cooling of micro/nano-mechanical resonators has been achieved via cavity cooling~\cite{Schliesser:cool:PRL:06, Groblacher:Nat:2009, Chan:Nat:2011} and active optical feedback cooling~\cite{Mancini:PRL:98} through radiation pressure~\cite{Mertz:APL:1993, Cohadon:PRL:99, Klechner:Nat:06,Li:NatPhys:2011} and photothermal~\cite{Barton:Nanolett:12, Usami:NPhot:2012} forces. A feedback system will suppress all motion of the oscillator and has no way to differentiate between motion due to thermal noise and a signal that one wishes to detect. The signal-to-noise ratio (SNR) of the sensor under conditions of steady-state feedback therefore may not be straightforwardly enhanced~\cite{Vitali:PRA:2002}. Not all is lost, however, as the increased effective damping rate due to feedback can give the oscillator a larger linewidth, which means that the oscillator will respond more quickly to external forces. As a result, less integration time is required to detect steady-state signals with the adoption of feedback cooling~\cite{Gavartin:NatNano:2012}.

Feedback cooling requires constant measurement of the system in order to generate a control signal to counteract thermalisation. If this measurement data is not used for feedback, one may ask if the data can still be used to improve our knowledge of the system dynamics and the measurement outcome. Provided that we are dealing with linear systems, this measurement data can indeed be fed into numerical estimation algorithms such as Kalman or Wiener filters. These techniques allow tracking and prediction of the motion, which can yield advantages similar to those of feedback cooling. In fact, it is possible to use numerical algorithms to simulate the effect of physical feedback cooling~\cite{Harris:PRL:2013}. These {\it estimation techniques} require accurate knowledge of the system parameters to perform efficiently. In a nano-scale opto-mechanical system, the phase and amplitude noise of the laser can alter the system parameters~\cite{Gavartin:Rep:NatNano:2013} and real-time tracking may be required for optimum estimation.

Outside the confines of the steady-state regime, feedback and estimation can be used to improve the SNR~\cite{Vitali:PRA:2002,Harris:PRL:2013}. To date, active cooling of nano-mechanical systems in the transient regime has not been observed and its implications for impulsive force sensing have not been studied. In the presence of an impulsive force, feedback cooling can be turned on prior to measurement and then turned off during the measurement. This {\it periodic quiescence feedback} works in such a way that the effect of an impulsive external force on the resonator is measured before the system is fully thermalized. In this case, the SNR of a pulsed signal can be improved by synchronizing the feedback in cycle with the signal. As with the steady-state conditions, it has also been predicted that an equivalent enhancement can be achieved by using estimation techniques provided the system dynamics are well known (see Supplementary Information of Ref.~\cite{Harris:PRL:2013}). 

In this letter we investigate the dynamics of gold-coated Ag$_{\textrm{2}}$Ga nanowires\footnote{http://nauganeedles.com}. Using homodyne detection we can measure the motion of the nanowire and then actively control the vibration of its modes using bolometric forces. Our setup is cavity-free, allowing cooling to be accomplished over a wider bandwidth where simultaneous cooling of multiple vibrational modes~\cite{VinantePRL:2008} is observed. We explore the implementation of non-stationary active cooling on nano-mechanical oscillators by means of optical forces and observe SNR enhancement of impulsive force measurements. We then compare the results with off-line estimation techniques and show enhancement by a factor of about $5$ using periodic feedback cooling as well as estimation methods. Using both physical control and estimation techniques we achieved a force sensitivity of better than $2 \times 10^{-16}$~\SI{}{N}.

\section*{Results}
\subsection{Detection and Feedback.}
\begin{figure}
	\centerline{\includegraphics[width=0.8\columnwidth]{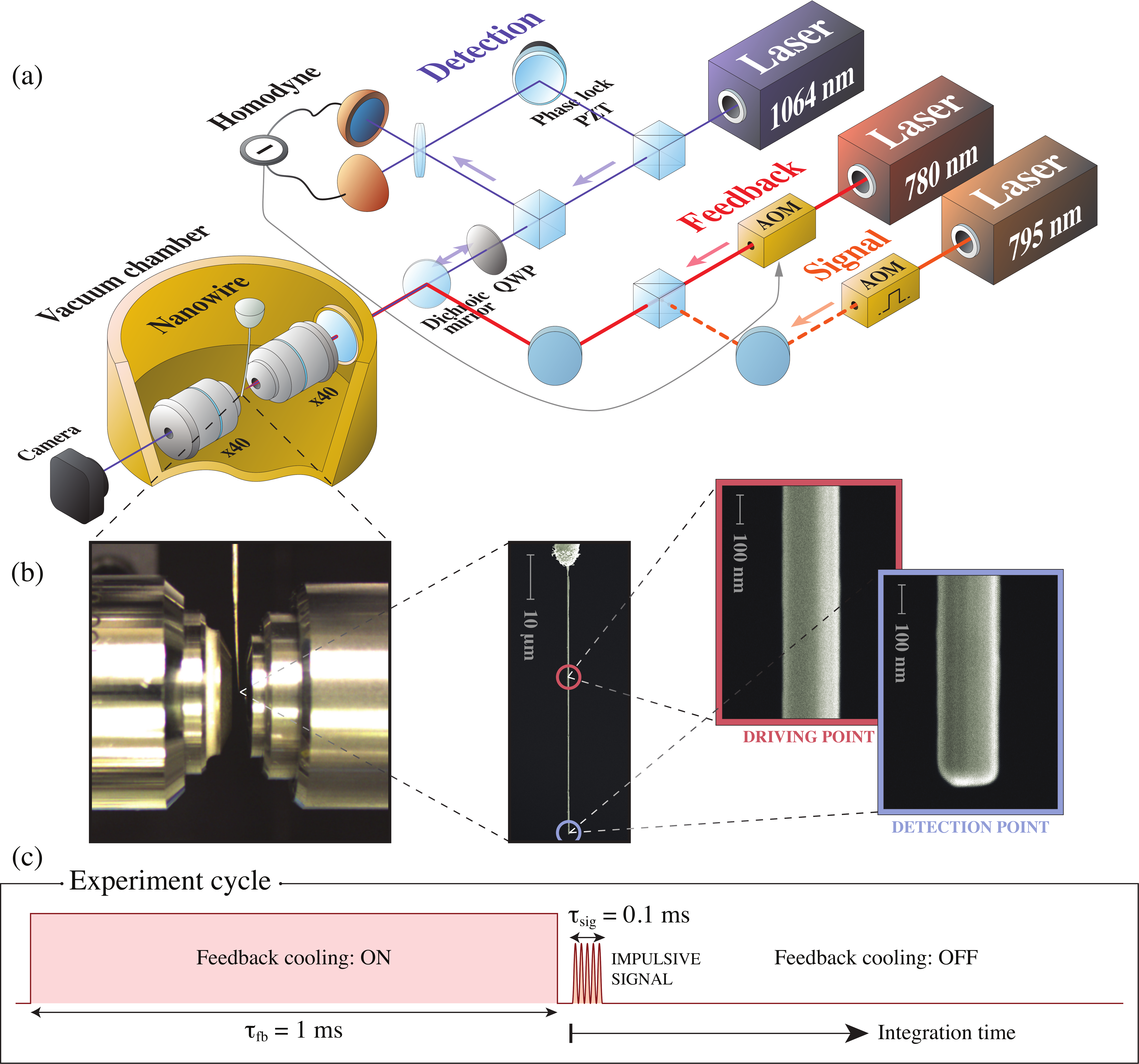}}
	\caption{{\bf Schematic experimental setup and experiment cycle.} \enspace(a) Experimental setup. The position of the nanowire is controlled by a nanopositioning stage. One laser (1064~nm) acts as a detection beam. Reflected and transmitted light from the nanowire are detected using a balanced homodyne and camera, respectively. Light is focused onto, and collected from, the nanowire using a pair of microscope objectives ($40\times$, $N\!A = 0.65$). A second laser (780~nm) is intensity modulated using an acousto-optic modulator, enabling feedback actuation based on the homodyne signal. During transient regime measurements, a third laser (795~nm) modulated to provide the impulsive signal is used. \enspace(b) Mounting arrangement and SEM close-ups of a cylindrical Ag$_{\textrm{2}}$Ga nanowire about \SI{60}{\micro m} long and \SI{150}{nm} thick. The detection beam interacts with the free end of the nanowire while the feedback or impulsive signal driving beams are typically positioned $\sim$\SI{10}{\micro m} from the tip to enhance the thermal bending. \enspace(c) Schematic of the experiment duty cycle during transient cooling and impulsive force measurement.}
	\label{fig: setup}
\end{figure}

The experimental setup is presented in Fig.~\ref{fig: setup}(a), with SEM images of a nanowire shown in Fig.~\ref{fig: setup}(b) (for more details on its properties and characterization refer to Supplementary Note 1). Optomechanical cooling often relies on optical cavity resonators to enhance the coupling to the mechanical modes. In that case the operational bandwidth is limited by the cavity dynamics and allows only a single mechanical frequency to be addressed via optical interaction.
In our experiment, we use a microscope objective to maximize the single-pass interaction of the nanowire with the laser light. The absence of a cavity allows simultaneous access to all of the nanowire's mechanical resonances. Light scattered from the nanowire is used in a homodyne measurement system~\cite{Mancini:PRL:98}. The motion of the nanowire is therefore referenced to the phase of the detection laser. The signal-to-noise ratio of this measurement is limited by the amount of scattering from the nanowire. The power used in the measurement is low enough to prevent the observation of any back-action due to the detection laser.

The actuation laser has a different wavelength to the detection laser to avoid any chance of interference between these subsystems. The dominant force on the nanowire is bolometric, i.e. it results from differential thermal expansion of a bimorph structure~\cite{Fu:APL:2011, Metzger:2004,Hossein-Zadeh:IEEE:2010}. In our case, it is the composite Au/Ag$_{\textrm{2}}$Ga structure of our nanowires that leads to strong thermal effects. The speed of the force depends on the thermal time constant of the object being heated, so it is naturally slow for macroscopic oscillators. For nanoscale objects, however, it can be both fast and substantial in magnitude~\cite{Dhara:PRB:2011, Pradhan:phd:2010}. The response time of the nanowire to laser driving is found to be \SI{15}{\micro s} at ambient pressure. Unlike radiation pressure, the direction of the force depends only on the structure of the object and is independent of the direction of the incident beam that is used to drive the nanowire. Experimental evidence of these aspects is offered in Supplementary Notes 2 and 3 and Supplementary Figures 1, 2, 3 and 4. Bolometrically actuated feedback cooling is somewhat counterintuitive, as the amplitude of the oscillations is subdued by use of a driving force of thermal origins - effectively cooling by heating. Even so, recent theoretical work shows that photothermal effects can assist cooling towards the quantum ground state \cite{Pinard:NJPh:2008, Restrepo:CRPh:2011,Abdi:2012gt} and photothermal back action in a cavity has also been used to cool a semiconductor membrane \cite{Usami:NPhot:2012} and a graphene sheet \cite{Barton:Nanolett:12}.

\begin{figure}
	\centerline{\includegraphics[width=0.8\columnwidth]{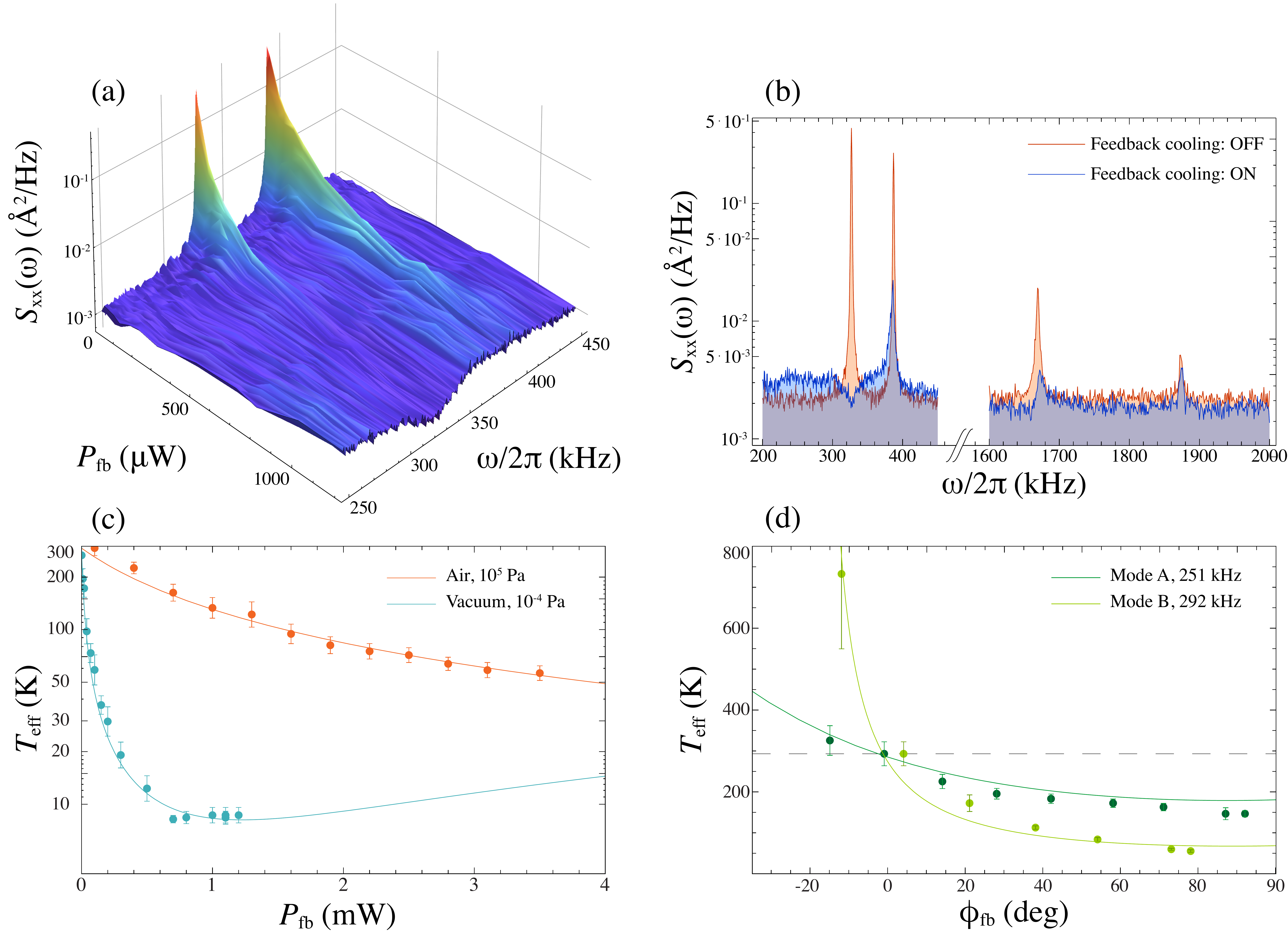}}
	\caption{{\bf Experimental results of multimode photothermal cooling of a nanowire.} (a) Displacement spectrum of the two orthogonal fundamental modes of a nanowire ($R \approx \SI{100}{nm}$ , $L \approx \SI{40}{\micro m}$) for different powers of the feedback beam at vacuum condition (\SI{E-4}{Pa}). \enspace(b) Simultaneous cooling of four distinguished resonances of the nanowire in vacuum with phase and gain optimized for more efficient cooling on one mode per pair. The displacement spectrum data in the absence of feedback is in red and the trace with feedback cooling is in blue. Insets show a close-up of the vibrational modes. \enspace(c) Effective temperature of the first vibrational mode as a function of feedback laser power at ambient (red) and vacuum (blue) conditions. The solid line represents the theory assuming a linear relation between laser power and feedback gain $g$. \enspace(d) Effective temperature of the first (dark green) and second (light green) vibrational modes as a function of the phase of the feedback signal. Phases of $-90^\circ$ and $90^\circ$ correspond to maximum oscillation amplification and maximum cooling, respectively. The solid lines represent the theory predictions. The error bars in (c) and (d) are estimated based on error propagation in the Lorentzian fit of the amplitude noise.}
	\label{fig: cooling}
\end{figure}

Fig.~\ref{fig: cooling}(a) shows the displacement spectrum of the nanowire for increasing power of the feedback beam in vacuum conditions. The phase of the feedback is adjusted using an adjustable bandpass analog filter. The feedback phase can be tuned such that simultaneous cooling of multiple modes of the nanowire can be obtained, as seen in Fig.~\ref{fig: cooling}(b) where we plot the uncooled (red) and cooled (blue) vibrational modes of the nanowire up to \SI{2}{MHz} for a particular phase and gain of the feedback loop. Applying a more advanced RF phase shifter, it is in fact feasible to efficiently cool vibrational modes over a broader frequency range. A comparison of the best effective temperatures reached in ambient and vacuum conditions is plotted as a function of feedback strength in Fig.~\ref{fig: cooling}(c); we have achieved cooling greater by an order of magnitude and for reduced driving power in vacuum compared to ambient pressure due to the improved mechanical quality factor. Fig.~\ref{fig: cooling}(d) shows the relation between phase and cooling for the first two oscillation modes of a nanowire in ambient pressure conditions; the difference in efficiency is attributed to dissimilarities in geometrical qualities along the directions of oscillation.

\subsection{SNR Enhancement.}
To measure the SNR in the transient regime, a \SI{0.1}{ms} long signal modulated at the mechanical frequency is sent to the nanowire right after the feedback is turned off (see Fig.~\ref{fig: setup}(a) and (c)). At the moment the signal arrives, with the feedback just switched off, the nanowire will also begin to thermalise at the rate of its mechanical dissipation, $\gamma_{\textrm{m}}$. Provided the signal time ($\tau_{\textrm{sig}}$) is much shorter then the re-thermalisation time ($\gamma_{\textrm{m}}^{-1}$), the measurement of the signal will not be significantly affected by thermal noise. In vacuum conditions the mechanical dissipation rate of the nanowire is lower than \SI{1}{kHz}, allowing integration times up to \SI{1}{ms} long. We integrate the energy of the signal after introducing the impulsive force and compare it with the thermal noise to estimate the SNR in the presence and absence of periodic cooling.

As we outlined earlier, linear feedback can also be simulated off-line using estimation techniques. We have performed both Kalman filtering and virtual feedback cooling (see {\it Methods}) on data taken in the absence of feedback. The digital Kalman filter is constructed based on prediction and update stages. At the prediction stage, the model of the system is used to estimate the evolution of the system for a short time into the future. In the update stage, the actual measurement results and known measurement and process noise vectors are used to refine the estimated evolution. The update stage of the filter is switched off when the impulsive force arrives, and the sizes of signal and noise are calculated as the phase-space distance between the measured and estimated trajectories in the presence and absence of the impulsive force. The virtual cooling method is used to directly simulate ideal periodic feedback cooling. A comparison of the estimation techniques with physical cooling in the transient regime is shown in Fig.~\ref{fig: snr}. Physical feedback cooling is shown to be as effective as the virtual cooling, which indicates near optimal actuation of the nanowire. For integration times longer than \SI{0.4}{ms}, feedback cooling shows slightly higher improvement than virtual cooling. This is likely due to perturbation of the system parameters by laser noise within the time scale of mechanical decay time. The Kalman filter is shown to outperform both physical and virtual cooling. In all cases, the SNR degrades with a rate corresponding to the mechanical decay time as the thermal noise contribution increases. The SNR enhancement factor $\eta_{\textrm{SNR}}$, defined as the ratio between the estimated SNR and that of the raw data where no cooling or filtering is used, is plotted in Fig.~\ref{fig: snr}(b). An enhancement factor of about $5$ has been observed for both the virtual cooling and Kalman filter methods.

We note here that while estimation methods can provide similar or even higher SNR enhancement factors than feedback cooling for short integration times, they require precise knowledge of system dynamics and parameters, and the production of the desired outcome can become computationally expensive. As an example we noticed that changing the mechanical frequency in the model by even $0.1\%$ could significantly change the SNR. Such a change in frequency can occur easily through a change in the bulk temperature of the oscillator.

To calculate the minimum force that can be measured with our setup, we consider the response $x(t)$ to a monochromatic force with magnitude $F_{\textrm{0}}$, duration $\tau_{\textrm{sig}}$ and frequency $\omega_{\textrm{0}}/2\pi$ that is $x(t) = \frac{F_{\textrm{0}}}{\sqrt{2\pi}} \int_{\textrm{0}}^{\tau_{\textrm{sig}}}\!dt' \chi(t-t')\sin(\omega_{\textrm{0}} t')$, where $\chi(t)$ is the mechanical susceptibility. The mean squared displacement (for integration time $\tau$) is $\left<x^2\right> =\frac{1}{\tau} \int_{\textrm{0}}^\tau\!dt \frac{F_{\textrm{0}}^2}{2\pi} \left|\int_{\textrm{0}}^{\tau_{\textrm{sig}}}\!dt' \chi(t-t')\sin(\omega_{\textrm{0}} t') \right|^2$, so, the force required to obtain a given mean squared displacement is
\begin{eqnarray}
F_{\textrm{0}} = \sqrt{\frac{2\pi \tau \left<x^2\right>}{\int_{\textrm{0}}^\tau\!dt \left| \int_{\textrm{0}}^{\tau_{\textrm{sig}}}\!dt' \chi(t-t')\sin(\omega_{\textrm{0}} t') \right|^2}}.
\end{eqnarray}
This suggests a sensitivity as low as \SI{2 e-16}{N} after integration time of less than \SI{0.1}{ms} for data filtered by the Kalman method. After \SI{0.1}{ms} the signal is switched off, so the relative contribution from the thermal noise increases and the sensitivity degrades. The high resolution and fast force sensing are important factors for bio-sensing where conformational changes typically happen on micro- to millisecond time scales~\cite{Dong:NNano:2009}.

\section*{Discussion}
In summary, we have demonstrated simultaneous active cooling of fundamental and higher-order modes of a bimetallic nanowire by means of optically induced thermal forces, obtaining temperatures as low as $8 \pm 1$~\SI{}{K}. In the transient regime, we investigated the effect of feedback cooling on the detection of an impulsive signal to demonstrate enhancement in measurement sensitivity. Although virtual feedback and estimation theories applied to off-line data provide similar improvements in sensitivity, they require that the system parameters are well known and constant, which may be a limitation in practical sensing applications. These technologies are increasingly important for sensing applications ranging from monitoring intermolecular interactions~\cite{Dong:NNano:2009} to subsurface imaging~\cite{Zhao:IOP:2010}, and generally provide an alternative to passive cooling for systems that can not be refrigerated. Both active feedback and numerical post-processing methods can be used to obtain an improvement in sensitivity over short time windows, especially advantageous when the system dynamics are changing rapidly and long integration times are not accessible~\cite{Craig:AustJChem:2006}.

\section*{Methods}
{\it Feedback cooling.} The power spectrum of the measured displacement of the nanowire driven by thermal forces in the presence of feedback is~\cite{Klechner:Nat:06}
\begin{eqnarray}
	S_{xx}(\omega)	=	\frac{2 \, \gamma_{\textrm{m}} \, k_{\textrm{B}} \, T_{\textrm{0}}}{m^2} \frac{1}{(\omega_{\textrm{m}}^2 - \omega^2)^2+(1+g)^2 \gamma_{\textrm{m}}^2 \omega^2},
	\label{eq:Sxx}
\end{eqnarray}
where $g$ is the effective gain that depends on the electronic gain, the mechanical quality factor, the thermal relaxation time and the induced mechanical rigidity resulting from photothermal forces~\cite{Metzger:2004}, $\omega_{\textrm{m}}/2\pi$ is the mechanical frequency, typically around \SI{300}{kHz}, $\gamma_{\textrm{m}}/2\pi$ is the mechanical linewidth, ranging from around \SI{10}{kHz} in air down to \SI{0.8}{kHz} in vacuum, $m$ is the nanowire's effective mass, around \SI{5}{fg}, $k_{\textrm{B}}$ is the Boltzmann constant and $T_{\textrm{0}}$ is the initial temperature.\\
The vibration of the nanowire introduces phase modulation on the reflected light. We send the DC homodyne signal to a piezo-actuated mirror to lock the phase of the local oscillator beam to that of the signal reflected from the nanowire. Above the locking bandwidth, which was around $10$~kHz, the error signal then becomes a readout of the nanowire motion. The effective temperature and mechanical damping rates when feedback is sent to the nanowire are given by $T_{\textrm{eff}} =T_{\textrm{0}}/(1+g)$ and $\gamma_{\textrm{eff}} = (1+g) \gamma_{\textrm{m}}$, respectively. The displacement spectrum has to be calibrated to compensate for the conversion and operational efficiency of the measurement devices. This is done by the integration of the area beneath the spectrum at room temperature and comparing it with the result expected from the equipartition theorem.\\
For higher gains a correction term~\cite{Poggio:PRL:2007} needs to be applied to the power spectrum (see Supplementary Note 4) and the effective temperature inferred. The latter is then characterized by a lower bound, $T_{\textrm{min}}=\sqrt{\frac{m \omega_{\textrm{m}}^2 \gamma_{\textrm{m}} T_{\textrm{0}}}{k_{\textrm{B}}}S_{\textrm{det}}}$, which depends on the spectral density of the detection noise $S_{\textrm{det}}$. The lower the measurement noise and the mechanical linewidth are, the lower the minimum temperature achievable will be. Once this limit is reached, increasing the gain further results in noise squashing~\cite{Buchler:OL:98,Poggio:PRL:2007} (see Supplementary Note 4 and Supplementary Figure 5).

\begin{figure}
	\centerline{\includegraphics[width=\columnwidth]{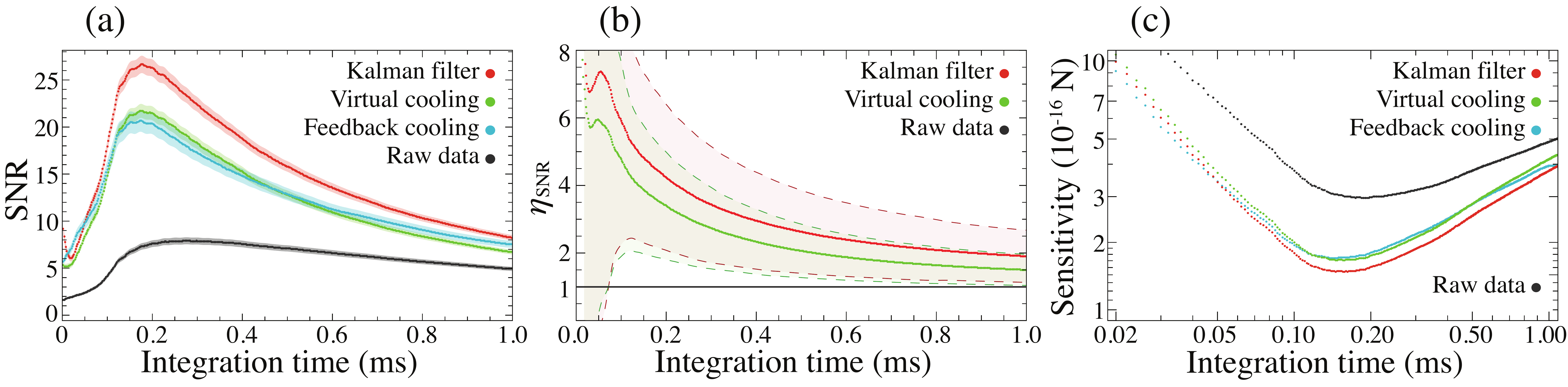}}
	\caption{{\bf Comparison of force measurement results of feedback cooling and estimation methods.} \enspace(a) Signal-to-noise ratio (SNR) measurement of energy of nanowire driven with impulsive photo-thermal force. The black and blue traces show the SNR measurement without and with feedback cooling. The red trace is the SNR of non-cooled data after post-processing using Kalman filter. The green trace represents the results of virtual cooling on data. The shaded regions represent the standard error. \enspace(b) The SNR enhancement factor, $\eta_{\textrm{SNR}}$, defined as the ratio of the estimated SNR divided by the SNR of the raw data. The dashed lines represent the standard deviation in enhancement estimation. (c) Force resolution as a function of integration time for filtered, raw and feedback cooling data.}
	\label{fig: snr}
\end{figure}

{\it Data processing.} To measure the SNR, four sets of homodyne signals are recorded at a rate of \SI{25}{MS~s^{-1}}, in the presence and absence of feedback and in the presence and absence of the impulsive force. We then apply a spectral filter to restrict the signal to a \SI{40}{kHz} bandwidth around the mechanical frequency. To calculate the SNR of, for example, feedback cooling data, we integrate the energy of the homodyne signal corresponding to the data taken in the presence of both periodic feedback and the impulsive force, and divide it by the average integral of the energy corresponding to data taken in the presence of periodic feedback and absence of the impulsive force. Analogous methods are used to measure SNR in the absence of feedback cooling, and when estimation techniques are used in place of physical cooling. All results are averaged over 150 traces, with the error calculated using standard error (for Fig.~\ref{fig: snr}(a)) or standard deviation (for Fig.~\ref{fig: snr}(b)).

{\it Estimation.} We use two different estimation strategies based on the discrete extended Kalman filter (EKF)~\cite{Iplikci:EKF2012} and the virtual cooling method~\cite{Harris:PRL:2013} to enhance the SNR (refer to Supplementary Note 5 for more details). To identify the state-transition matrix of the EKF, we need to know the natural frequency, damping rate, initial amplitude, initial velocity, time interval, process and measurement noise vectors and initial covariance estimates. In order to propagate the state and estimate the evolution of the system, we use the Runge-Kutta (or RK4) method to predict the state of the system in the next step and update the system's quadratures according to the relative uncertainties of measured and predicted values. We switch off the updating step of the EKF after \SI{1}{ms}, at the time when the impulsive force arrives, so that the EKF predicts the subsequent behaviour of the system. This prediction is expected to be reliable while the system is not significantly affected by stochastic thermal noise, and deviations from it indicate the presence of the impulsive signal. The SNR is calculated in the same way as for feedback cooling, but instead of integrating the energy of the oscillator we integrate the phase-space distance between the observed and predicted trajectories.\\
The virtual cooling is applied as suggested in the Supplementary Information of Ref.~\cite{Harris:PRL:2013} in order to simulate the periodic feedback cooling scheme. Knowing the mechanical susceptibility of the system we perform temporal discretization of the resulting Fredholm equation into a $1000\times 1000$ matrix equation. Solving this yields the simulated measurement record, from which the SNR is calculated (refer to Supplementary Note 5 for more details).

{\bf Acknowledgements}\\
The authors would like to thank W.\ P.\ Bowen, A.\ E.\ Miroshnichenko and A. Mirzaei for fruitful discussions and Nauganeedles' team for their cooperation. This research was conducted by the Australian Research Council Centre of Excellence for Quantum Computation and Communication Technology (project number CE110001027).

{\bf Author contributions}\\
M.H.\ and G.G.\ designed and conceived the experiment; H.J.S., M.H.\ and G.G.\ performed modelling and data analysis; M.H.\ and G.G.\ prepared the manuscripts with critical comments from other authors. All stages of the work were supervised by B.C.B.\ and P.K.L.

{\bf Additional information}\\
The authors declare no competing financial interests. Supplementary information accompanies this paper at www.nature.com/ncomms. Reprints and permission information is available online at http://www.nature.com/reprints. Correspondence and requests for materials should be addressed to B.C.B.\ or P.K.L.


\clearpage
\section*{Supplementary Notes}

\subsection{Supplementary Note 1 - Characterization.}
The oscillators used in our system are gold-coated Ag$_{\textrm{2}}$Ga nanowires (provided by NaugaNeedles~LLC~\cite{sup:nauganeedles}). They range in size between \SI{20} and \SI{60}{\micro m} in length and \SI{50} to \SI{200}{nm} in diameter, and are coated with about \SI{40}{nm} of gold to improve scattering efficiency, as optical detection of its vibrational modes relies entirely on light scattered from its surface. The scattering efficiency ($Q_{\textrm{sca}}$) of a long cylinder for light normally incident on the nanowire indicates the amount of light scattered from the sub-wavelength object predicted by Mie scattering theory~\cite{sup:BohrenHuffman1983}. This quantity is plotted as a function of nanowire radius in Suppl.\ Fig.~\ref{fig: scattering}(a), with the total scattering efficiency in the \textsc{xy}-plane around the nanowire placed at the origin, when laser light is incident from right to left, shown in the inset. It can be used to calculate the radiation pressure (or scattering) force, $F_{\textrm{sca}}(r) = 4 Q_{\textrm{sca}} w R I(r)/c$, and absorption force, $F_{\textrm{abs}}(r) = 4 Q_{\textrm{abs}} w R I(r)/c$, exerted on cylindrical nanowires, where $w$ is the beam waist, $R$ the radius of the nanowire, and $I(r)$ is the intensity distribution of the incident laser light along the width of the nanowire. $Q_{\textrm{abs}}$ is the absorption coefficient calculated using scattering and extinction efficiencies~\cite{sup:Biedermann2009}. These forces are plotted in Suppl.\ Fig.~\ref{fig: scattering}(b). It is evident that the absorption force is much weaker than the scattering forces. Experimentally, we have observed that the bolometric force $F_{\textrm{bol}}$ is the dominant photo-induced force on the nanowire, as the Au$/$Ag$_{\textrm{2}}$Ga bimorph material responds very efficiently to optically induced thermal bending. In our system, we estimate the bolometric force to be around 100 times bigger than the radiation pressure force.

The main detection method involves the use of scattered light in an interferometric manner. Displacement of the nanowire along the optical axis modulates the phase of the scattered light. The reflection interferes with a local oscillator in a homodyne measurement setup, providing a way to detect the phase modulation. Using a PID controller, we lock the phase of the homodyne system to the phase quadrature of the scattered light. Outside the control bandwidth, the motion of the nanowire is then proportional to the error signal of the control loop. Light interacting with the nanowire and transmitted through the microscope lenses is also used for detection with a split detector. This method detects the motion of the resonator on the direction transverse to the optical axis and is complementary to the interferometric measurement. A juxtaposition of the two methods is presented in Suppl.\ Fig.~\ref{fig: scattering}(c). Here we see two modes of vibration at different frequencies.  In fact these peaks are due to spatially orthogonal modes of the nanowire.

The homodyne measurement efficiency of a vibrational mode depends on the angle between the direction of oscillation and the axis of detection: if the oscillation is perfectly aligned with the optical axis, for example, that mode will be fully detected by the homodyne system and vibrations that are spatially orthogonal will not be observed.  An example of this situation is shown in Suppl.\ Fig.~\ref{fig: Angle}(a) where we have rotated the nanowire until only one mode is visible. When the detection beam is incident with a \SI{45}{\degree} angle relative to the two orthogonal modes, both can be detected.  This is the case in Suppl.\ Fig.~\ref{fig: scattering}(c) where both the split detector and the homodyne system can see the motion of the nanowire.  The angle in this case is not exactly \SI{45}{\degree} since the peak heights are not identical for both methods, rather the largest peak for the homodyne signal is the smallest in the split detector signal, and vice-versa, as we expect.
 

\subsection{Supplementary Note 2 - Deflection.}
When subject to feedback from an intensity-modulated beam, the raise in bulk temperature of the nanowire, taking into account the reflectivity of the gold layer~\cite{sup:Oloomi:Au2010} and an amplitude modulation depth of $0.1\%$ for a \SI{1}{mW} laser beam, is estimated to be around \SI{10}{K}. It has been demonstrated that such temperature increase can cause thermally-induced nanomechanical deflection of a few nm of hybrid nanowires~\cite{sup:IkunoAPL:2005}.\\
We were able to estimate the deflection by systematic modulation of the driving field for different powers. By turning the field on and off at low frequency to produce a variation in the detection signal, and by calibrating the amplitude of these variations to the total amplitude of the interference fringes of the unlocked homodyne signal (equivalent to one wavelength), we were able to measure the thermal time response (Suppl.\ Fig.~\ref{fig: Deflection}(a)-(b)) and observed deflections of about \SI{60}{nm} with less than \SI{5}{mW} of power (Suppl.\ Fig.~\ref{fig: Deflection}(c)). These results reveal that a minimum power is required to drive the nanowire.\\
The sign of the feedback gain can determine whether both modes can be cooled or heated simultaneously. If the incident beam (or, equivalently, the nanowire) is rotated by \SI{90}{\degree}, the relative phase between one of the modes and feedback will be opposite, while the other will be the same as before. This will produce cooling of one mode and heating of the other, as shown in Suppl.\ Fig.~\ref{fig: Angle}(b). A rotation of a further \SI{90}{\degree} will again sync the phase of the two modes relative to the feedback, although now the phase that was originally used for cooling will produce heating, and vice versa. The full-rotation periodicity is a consequence of the thermal nature of the force inducing directional bending, which depends purely on geometrical properties and not on the direction of the incident light: if that were to be the case, the behaviour after a \SI{180}{\degree} rotation should be the same as that of the nanowire in the original position, as the movement caused by radiation pressure forces is always in-phase with the driving field.

\subsection{Supplementary Note 3 - Feedback cooling.}
In general, the equation of motion~\cite{sup:Metzger:2004} of the nanowire is given by
\begin{eqnarray}
	 \ddot{x} + \gamma_{\textrm{m}} \dot{x} + \omega_{\textrm{m}}^2 x = F_{\textrm{th}} + \sum_n{\int_{\textrm{0}}^t{\dot{F}_n(t') h_n(t-t')dt'}},
\end{eqnarray}
where $\omega_{\textrm{m}}$ is the mechanical frequency, $\gamma_{\textrm{m}}$ is the mechanical damping, $F_{\textrm{th}}$ is the thermal force driving the Brownian motion, and $F_n$ includes all of the other forces such as radiation pressure and photothermal forces. The term $h_n(t) = 1-e^{-t/\tau_{\textrm{c}}}$ denotes the response time of the structure for different forces with time constant $\tau_{\textrm{c}}$. In the case of the radiation pressure force this is almost instantaneous ($h_{\textrm{rp}}(t) \simeq 1$), but response time of photothermal forces depends on the thermal conductivity of the material and geometry of the structure. The response time of nanocylinders to temperature variations depends on radius and thermal diffusivity $\kappa$ as $\tau_{\textrm{c}} \simeq R^2/(4 \kappa)$~\cite{sup:Anderson:Sci:1983}. Material properties and size strongly affect the thermal diffusivity, and for nano-scale objects it can be almost 2-3 orders of magnitude smaller compared to the bulk material~\cite{sup:Dhara:PRB:2011, sup:Pradhan:phd:2010} due to phonon scattering overcoming the phonon-phonon coupling. Assuming an effective thermal diffusivity of approximately \SI{e-6}{m^2 s^{-1}}, the thermal relaxation time of the nanowire can be on the order of ns or smaller, implying a response faster than its mechanical vibrations. The heat generated by the laser power will diffuse inside the nanowire more rapidly than the modulation, allowing the nanowire to react quickly to temperature fluctuations.\\
To explain some of the observed effects, such as the relative behaviour between orthogonal modes, and to be able to predict and have direct control on the system, we developed a simple model for our feedback setup that takes into account the transfer function of the system. The model includes a stability analysis of the feedback loop, from which it emerges that in our parameter regime we are far from any singularity. Thanks to the comparison of the total transfer function for both phase and gain with the experimental data collected we gained valuable insight on the spatial orientation of the modes. Suppl.\ Fig.~\ref{fig: FeedbackTheory} shows the experimental transfer functions and simulated feedback responses for the nanowire at two different orientations. The experimental transfer function is used to infer the orientation angle of the nanowire. Using this information, the model predicts accurately the behaviour of the two orthogonal modes and their relative detection.

\subsection{Supplementary Note 4 - Noise squashing.}
For high gains, the measurement noise becomes significant and it is important to include a correction term in the expression for the displacement power spectrum observed~\cite{sup:Poggio:PRL:2007},
\begin{eqnarray}
	S_{xx}(\omega)		=	\frac{2 \, \gamma_{\textrm{m}} \, k_{\textrm{B}} \, T_{\textrm{0}}}{m^2} \frac{1}{(\omega_{\textrm{m}}^2 - \omega^2)^2+(1+g)^2 \gamma_{\textrm{m}}^2 \omega^2} + S_{\textrm{det}}\frac{(\omega_{\textrm{m}}^2 - \omega^2)^2 + \gamma_{\textrm{m}}^2 \omega^2}{(\omega_{\textrm{m}}^2 - \omega^2)^2+(1+g)^2 \gamma_{\textrm{m}}^2 \omega^2},
	\label{eq:Sxx}
\end{eqnarray}
which depends on the spectral density of detection noise, $S_{\textrm{det}}$. The temperature inferred from this spectrum is
 \begin{eqnarray}
 	T_{\textrm{eff}} = \frac{T_{\textrm{0}}}{1+g} + \frac{g^2}{1+g} \frac{m \omega_{\textrm{m}}^2 \gamma_{\textrm{m}}}{4 k_{\textrm{B}}} S_{\textrm{det}},
\end{eqnarray}
presenting now a lower bound $T_{\textrm{min}}=\sqrt{\frac{m \omega_{\textrm{m}}^2 \gamma_{\textrm{m}} T_{\textrm{0}}}{k_{\textrm{B}}}S_{\textrm{det}}}$.\\
A characteristic of these correction terms is that the displacement spectrum can get lower than the shot noise level. This is in fact typical of feedback systems, since the noise measured by our main detection scheme is inside the feedback loop and is not a faithful representation of the real noise of the system. The random fluctuations on the homodyne local oscillator beam are part of the information sent through the feedback to the nanowire, and are transferred to the phase of the homodyne detection beam once it is scattered back by the oscillator. When the scattered light interferes with the local oscillator, the noise on the two beams is now correlated and can destructively interfere. As a result, the measured noise can appear \emph{squashed} below the shot noise level~\cite{sup:Buchler:OL:98}; out-of-loop detection would reveal this to be a property inherent to the feedback. Other experiments have also observed this phenomenon~\cite{sup:Poggio:PRL:2007}.
To deliberately observe noise squashing, we increased the local oscillator power to have larger shot noise. As it can be seen in Suppl.\ Fig.~\ref{fig: Squashing}(a), applying the feedback with significant gain results in noise squashing where the amplitude noise measured by homodyne goes below the shot noise. The minimum temperature $T_{\textrm{min}}$ is reached when the Lorentzian profile lies flat on the noise level; for higher gain, squashing becomes manifest and the effective temperature gets higher again. An example of inferred temperature for increasing powers of the cooling beam (i.e.\, increasing $g$) is shown in Suppl.\ Fig.~\ref{fig: Squashing}(b), where the minimum temperature is reached at $\sim \SI{0.5}{mW}$ of power.

\subsection{Supplementary Note 5 - Estimation techniques.}
Given the dynamics of some system, the (extended) Kalman filter~\cite{sup:TheKalman} processes a series of noisy measurements of the system in order to keep track of a statistically optimal estimate of its underlying state. Our implementation of the extended Kalman filter was based on the code at reference~\cite{sup:Casper:Kalman}. Since the specifics of the filter are quite involved we will not discuss them in full detail here, and only indicate the general procedure. For a more in-depth discussion we refer the reader to the references~\cite{sup:Casper:Kalman} (for implementation details) and \cite{sup:Sorenson:Kalman, sup:Maybeck:Kalman, sup:WelchBishop:Kalman} (for complete introductory discussions).

In our implementation we store the system state as a 4-vector containing the oscillator's position $x$, velocity $\dot x$, damping ratio $2\zeta=\gamma_{\textrm{m}}/\omega_{\textrm{m}}$ and mechanical frequency $\omega_{\textrm{m}}$. We also keep track of a covariance matrix which describes the uncertainty in this state vector. At each timestep (when a new measurement is taken, which is every \SI{2}{\micro \second} in our case), the filter updates in two stages. In the first ``prediction'' stage, the initial value problem describing the evolution of the system from the current state is solved using the Runge-Kutta method (RK4) to predict the state at the next timestep. In the RK4 approximation, the next value is determined by the present value plus the weighted average of four increments, where each increment is the product of the time interval $dt=\SI{2}{\micro \second}$ with the derivative of the vector $\mathbf F(t) = ( x(t), \dot{x}(t), \zeta, \omega_{\textrm{m}} )$ (calculated from the equation of motion for the system).

In the second ``updating'' stage, the predicted oscillator position is compared to the measured value, and the estimated state vector is updated to reduce this difference in accordance with the relative uncertainties (that is, more weight is given to either the measurement or the estimated state depending on which quantity has the lower uncertainty). The resulting state vector estimate is used as the initial state for the next timestep, and so on. This procedure yields an estimated trajectory of the oscillator, and from this we may extract a signal as described in the main text.

To apply this algorithm we needed to define initial values for the mechanical frequency $\omega_{\textrm{m}}/2\pi$ and linewidth $\gamma_{\textrm{m}}/2\pi$. These were first roughly estimated from the raw data, and then repeatedly adjusted in order to maximise the resulting SNR of the filtered trajectories. The final values we used were $\omega_{\textrm{m}}/2\pi = \SI{339.9}{kHz}$ and $\gamma_{\textrm{m}}/2\pi=\SI{0.53}{kHz}$. We note that this linewidth is smaller than that estimated from the raw data (roughly $\SI{0.8}{kHz}$). It is not entirely clear why it is advantageous (in terms of extracting the SNR) to model the oscillator with a longer thermal relaxation time, but a possible explanation is obtained by considering the behaviour of the Kalman filter's prediction algorithm. In particular, we recall that to determine the SNR we turn off the updating stage of the filter after \SI{1}{\milli\second}, and quantify the signal by comparing the predicted trajectory to the observed trajectory. The predicted trajectory is simply a sinusoid decaying with the thermal relaxation time, whereas in reality these oscillations do not decay since the nanowire is always being driven by thermal noise (even when no signal is applied). A longer thermal relaxation time in the model causes the predicted oscillations to decay less slowly and thus, on average, provide a better prediction of the true oscillations.

We implemented the virtual cooling method as described in reference \cite{sup:Harris:PRL:2013}. The idea is that the raw measurement record of the oscillator's position (taken in the absence of feedback) may be post-processed to yield the measurement record that would have been obtained if a particular feedback scheme had been employed. We assume an ideal feedback force which is switched off at time $t_{\textrm{0}}$,
\begin{equation}
F(t,x):=gH(t_{\textrm{0}}-t)x(t)=\int_{\textrm{0}}^t \delta(t-\tau)gH(t_{\textrm{0}}-t)x(\tau)d\tau,
\end{equation}
where $x$ is the measurement record, $g$ is the feedback gain, $H$ is the Heaviside step function and $\delta$ is the Dirac delta function. As per \cite{sup:Harris:PRL:2013}, we then define a function
\begin{equation}
 h(t,\tau):=\int_{-\infty}^\infty \delta(t'-\tau)g H(t_{\textrm{0}}-t')\chi(t-t')dt'=g H(t_{\textrm{0}}-\tau)\chi(t-\tau),
\end{equation}
where $\chi$ is the mechanical susceptibility. The function $h$ describes the response of the system at time $t$ to feedback resulting from the position of the oscillator at time $\tau$, and is used to simulate the feedback process. Specifically, it is shown in reference \cite{sup:Harris:PRL:2013} that the simulated measurement record $x_*$ satisfies the Fredholm equation
\begin{equation}
x_*(t) - \int_{\textrm{0}}^\infty d\tau h(t,\tau) x_*(\tau) = x(t).
\label{sim_equation}
\end{equation}
To solve this equation we again follow the procedure of reference \cite{sup:Harris:PRL:2013} and discretize to $1000$ timesteps of length $2\mu$s, so that $h$ is represented by a $1000 \times 1000$ matrix $\mathbb H$, while $x$ and $x_*$ are represented by vectors $\mathbf x$ and $\mathbf x_*$ of length $1000$. Equation~\ref{sim_equation} becomes $\mathbf x_*-\mathbb H\mathbf x_*=\mathbf x$, which is solved to obtain the simulated measurement record $\mathbf x_* = (\mathbb I-\mathbb H)^{-1}\mathbf x$ (where $\mathbb I$ is the $1000 \times 1000$ identity matrix).

As with the Kalman filter, the virtual cooling method required several initial parameters to be defined. We determined these as follows. First, rough values of mechanical frequency $\omega_{\textrm{m}}/2\pi$, linewidth $\gamma_{\textrm{m}}/2\pi$ and time $t_{\textrm{0}}$ were estimated from the raw data, and then the gain $g$ was varied to maximise the SNR of the resulting simulated measurement record. The first three parameters were then repeatedly adjusted in order to maximise the peak SNR obtainable by varying the gain. The final values used were $\omega_{\textrm{m}}/2\pi=\SI{339.722}{kHz}$, $\gamma_{\textrm{m}}/2\pi=\SI{0.85}{kHz}$, $t_{\textrm{0}}=\SI{0.896}{ms}$ and $g=-1.6784\times\exp(-0.00004i)\times 10^{-9}$.

These two estimation techniques allow a raw trace to be filtered in order to enhance the SNR of any signal present in the oscillations. We denote by $\eta_{\textrm{SNR}}$ this enhancement factor (or rather, the average of enhancement factors over all traces). Note that it does not make sense to discuss this factor in relation to feedback cooling, since if we are given a cooled trace we cannot ``undo'' the cooling to calculate the SNR that we would have obtained without cooling. Thus for feedback cooling we may define the enhancement in the \emph{average} SNR (by comparing the average SNR with feedback cooling to the average SNR without), but not the trace-by-trace enhancement factor $\eta_{\textrm{SNR}}$, therefore it is not reasonable to estimate the standard deviation of enhancement factor for feedback data.

\section*{Supplementary References}


\clearpage
\section*{Supplementary Figures}
\setcounter{figure}{0}
\renewcommand{\figurename}{Supplementary Figure}

\begin{figure}[!h]
	\centerline{\includegraphics[width=\columnwidth]{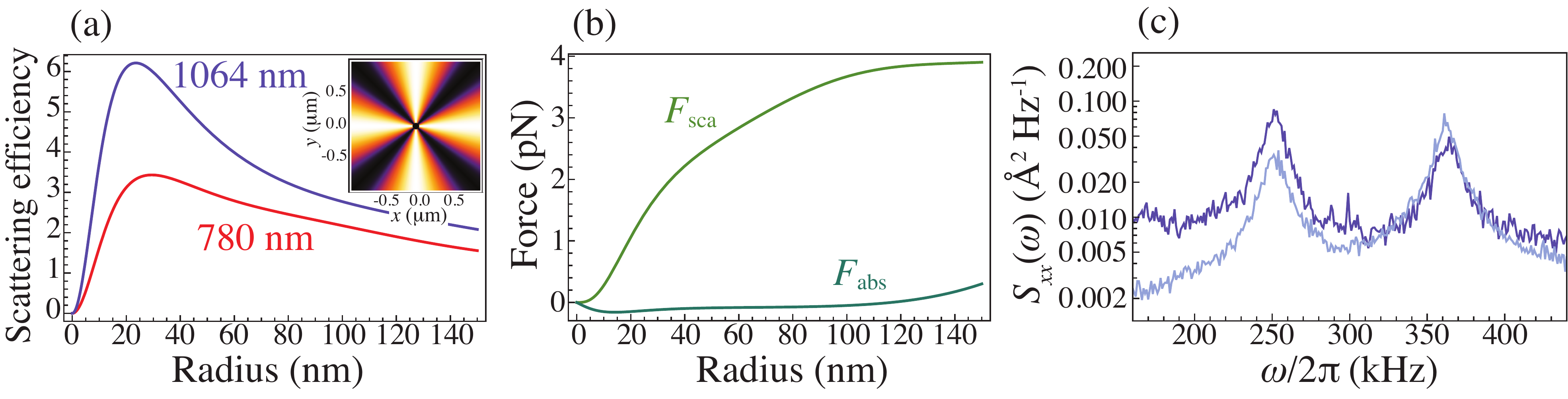}}
	\caption{{\bf Scattering simulation and detection.} \enspace(a) Total nanowire scattering efficiency as a function of its radius for two different wavelengths, $\lambda = \SI{780}{nm}$ and \SI{1064}{nm}. Inset shows the total scattering efficiency in the \textsc{xy}-plane (seen from below) around the nanowire placed at the origin, when laser light is incident from right to left. \enspace(b) Absorption and scattering forces exerted on the nanowire as a function of its radius for $\lambda = \SI{1064}{nm}$ and input power of 1 mW. \enspace(c) Detection of two orthogonal modes of the nanowire, using interferometry of the reflected beam (dark) and split detection of the transmitted beam (light). The two peaks are detected with different efficiency depending on the angle of the mode with the axis of detection, parallel to the optical axis for reflection and perpendicular to it for transmission.}
	\label{fig: scattering}
\end{figure}

\begin{figure}
	\centerline{\includegraphics[width=\columnwidth]{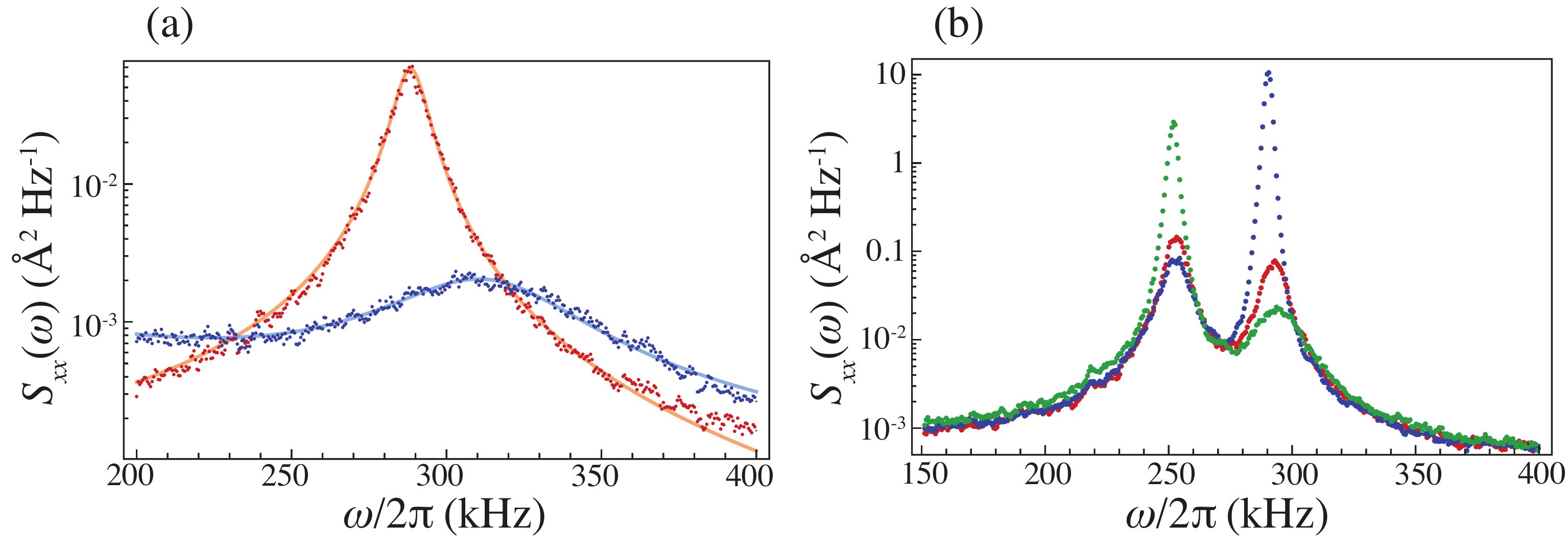}}
	\caption{{\bf Effect of detection angle and phase on cooling.} \enspace(a) The displacement spectrum of the nanowire used for cooling as shown in the main article, with (blue) and without (red) the feedback signal. The nanowire has been rotated so that only a single vibrational mode is visible in this frequency range. Note that this is \SI{45}{\degree} rotated compared to the situation for the data shown in Fig.~2 of the main manuscript. The solid lines represent the theory. \enspace(b) The displacement spectrum of the two modes of the same nanowire rotated by a further \SI{45}{\degree} relative to its orientation in (a). By swapping the sign of the gain, the behaviour of the feedback is altered in a mode selective manner.  The red trace is obtained without feedback, blue and green data correspond to feedback with $0$ and $\pi$ phase shift, respectively.}
	\label{fig: Angle}
\end{figure}

\begin{figure}[!h]
	\centerline{\includegraphics[width=\columnwidth]{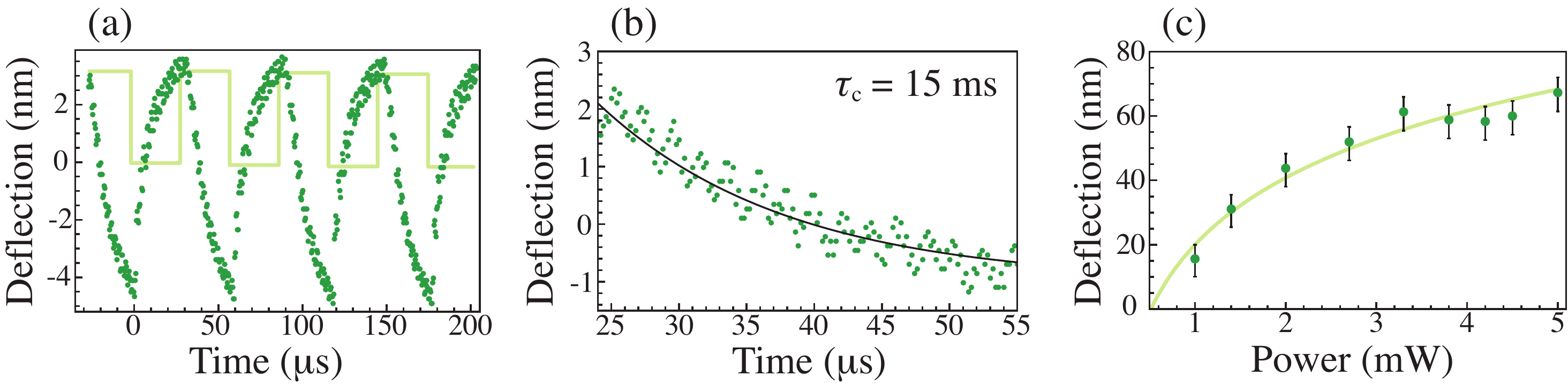}}
	\caption{{\bf Characterization of the nanowire deflection.} \enspace(a) Deflection of the nanowire measured using the homodyne signal while power is modulated at a rate of 17 kHz (shown as a square wave). \enspace(b) Exponential fit of deflection amplitude decaying after the driving force is switched off. The high-frequency oscillations on top of the exponential decay represent oscillations at the mechanical frequency. \enspace(c) The quasi-static displacement measured at different laser powers. The line is a guide to the eye. We infer a bolometric force of about \SI{e-10}{N} at \SI{1}{mW} of power, taking into account a spring constant of \SI{2 e-2}{N m^{-1}}. }
	\label{fig: Deflection}
\end{figure}

\begin{figure}[!h]
	\centerline{\includegraphics[width=0.9\columnwidth]{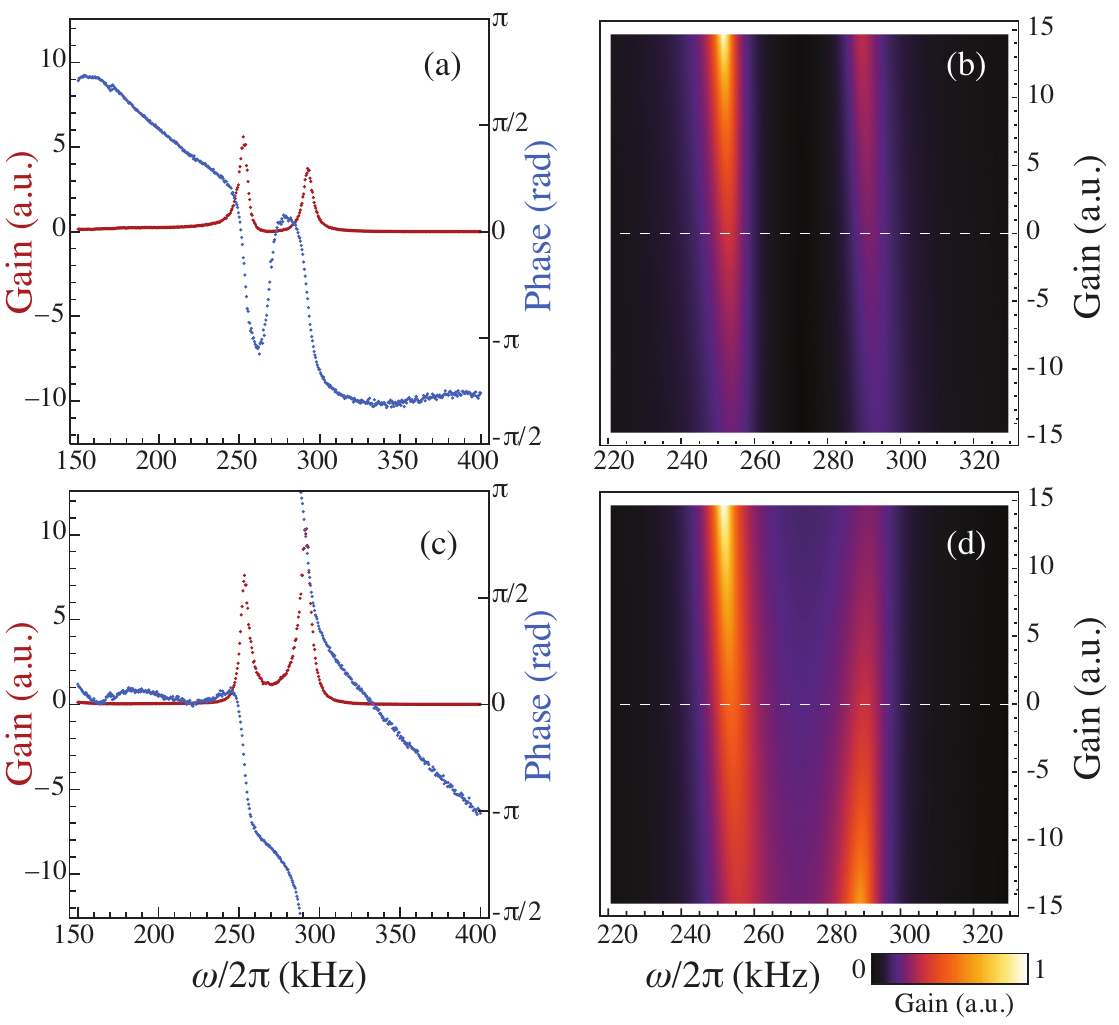}}
	\caption{{\bf Transfer function and feedback modelling for opposite orientations.} \enspace(a) Experimental transfer function of gain (green) and phase (red) for the nanowire in the orientation used for the data shown in the main article.\enspace(b) Feedback model prediction for the transfer function at different gain values. Increasing gain causes heating of both modes, while gains at opposite phase lead to cooling as shown in the main text. The dashed line indicates absence of feedback. \enspace(c) Similar to (a), for the nanowire rotated by \SI{90}{\degree}. \enspace(d) Feedback model prediction for the new orientation inferred by the transfer function. The two modes are out of sync, and one is heated while the other one is cooled. For comparison, Suppl.\ Fig.~\ref{fig: Angle}(b) shows the experimental power spectrum corresponding to this orientation.}
	\label{fig: FeedbackTheory}
\end{figure}

\begin{figure}[!h]
	\centerline{\includegraphics[width=\columnwidth]{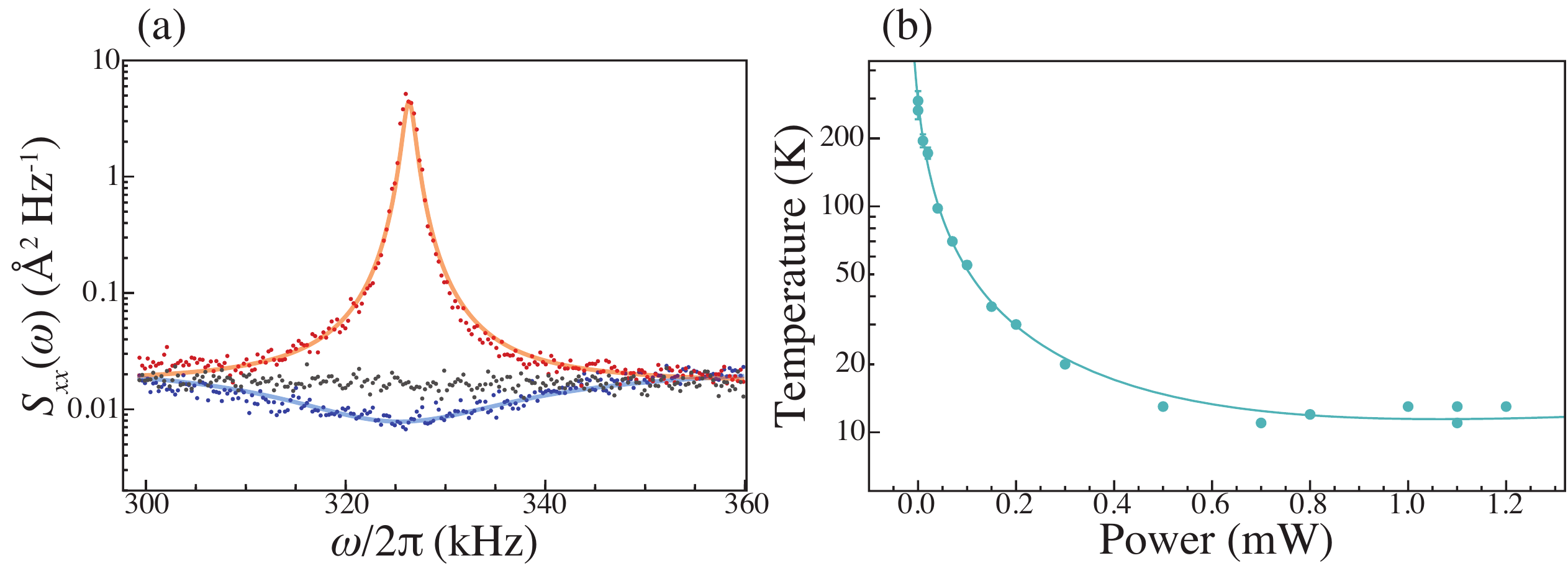}}
	\caption{{\bf Noise squashing and feedback cooling at higher shot noise levels.} \enspace(a) Noise spectrum showing the thermal amplitude noise of the nanowire (red), shot noise (black) and feedback result (blue) representing the noise squashing. \enspace(b) Temperature of the first vibrational mode of the nanowire as a function of power of the cooling beam. The solid lines in (a) and (b) are theory fits.}
	\label{fig: Squashing}
\end{figure}

\end{document}